\documentclass[useAMS,usenatbib]{mn2e}

\usepackage{graphicx}
\usepackage{amssymb}
\usepackage{latexsym}
\usepackage{amsmath}
\usepackage{txfonts}
\usepackage{natbib}

\title[Pupil plane optimization for multiaxial interferometry]{Pupil plane optimization for single-mode multiaxial optical interferometry with a large number of telescopes}

\author[J.-B. LeBouquin and E. Tatulli]{J.-B. LeBouquin$^{1,3}$\thanks{E-mail: jlebouqu@eso.org} and E. Tatulli$^{2,3}$\\
$^{1}$ European Southern Observatory, Casilla 19001, Santiago 19, Chile\\
$^{2}$ Osservatorio di Arcetri, L.go E. Fermi, 5, 50125 Firenze, Italia\\
$^{3}$ Laboratoire d'Astrophysique, Observatoire de Grenoble, 38041 Grenoble, France}

\begin{document}

\date{Accepted 2006 July 12. Received 2006 July 7; in original form 2006 May 29}

\pagerange{\pageref{firstpage}--\pageref{lastpage}} \pubyear{2006}
\maketitle
\label{firstpage}

\begin{abstract}
Incoming and planed optical long baseline interferometers will allow rapid spectro-imaging at high angular resolution. Non-homothetic Fizeau instrument using optical fibers is one of the most promising concept because it combines good sensitivity and high spectral resolution capabilities. However, when increasing the number of input telescopes, one critical issue is the design of the beam recombination scheme, at the heart of the instrument. 
Extending our previous analysis on the multiaxial "all-in-one" recombination, where the beams are mixed all together \citep{Tatulli-2005}, we tackle in this paper the possibility of reducing the number of pixels that are coding the fringes by compressing the pupil plane from partially redundant output pupils configuration. Shrinking the number of pixels -- which drastically increases with the number of recombined telescopes -- is indeed a key issue that enables to reach higher limiting magnitude, but also allows to lower the required spectral resolution and fasten the fringes reading process.
By means of numerical simulations, we study the performances of existing estimators of the squared visibility with respect to the compression process.
We show that, not only the model based estimator lead to better signal to noise ratio (SNR) performances than the Fourier ones, but above all it is the only one which prevent from introducing baseline mixing biases in the visibilities as the pupil plane compression rate increases. Furthermore, we show that moderate compression allows to keep the visibilities SNR unaffected. In the light of these conclusions, we propose an optimized pupil arrangements for 6 and 8 beam recombiners.
\end{abstract}



\section{Introduction} \label{sec:intro}
After the first results of COAST \citep{Baldwin-1996, Young-2000jul}, NPOI \citep{Hummel-1998jul} and IOTA \citep{Monnier-2004b}, the next challenge of optical long baseline interferometry is to commonly perform spectro-imaging of faint sources. From the end of 2005, this technique has moved one step forward with the operating of the AMBER instrument \citep{Petrov-2003spie}, the near-infrared combiner of the Very Large Telescope Interferometer \citep[VLTI,\ ][]{Glindemann-2003}. However, with its 3 beams, it will require several nights to be able to restore consistent images \citep{Thiebaut-2003}. Then, huge improvements are contemplated to be accomplished with second-generation instruments that will use $4$, $6$ or even $8$ telescopes coupled with spectral resolution abilities \citep{Malbet-2004}. One critical point in the design is the choice of the beam recombination concept, heart of the instrument.

The interferometric observables are the complex coherence factors (amplitude and phase) of the fringes formed by each pair of beams (the so-called baselines). They contain the information related to the spatial distribution of the source at high angular resolution. The simplest way to recover all the available information is to mix all the beams together (all-in-one scheme), as their is no need to split and rearrange the beams as in a pairwise design. Besides, it leads to better performances in the photon noise regime, because \emph{all photons} are used to create \emph{all fringes}. Those fringes appear by modulating the optical path differences between the beams. That can be done temporally (coaxial or Michelson modulation) or spatially (multiaxial or Fizeau modulation). Between these two solutions, recent studies emphasized the advantages of multiaxial concept thanks to less beam-splitters, mirrors and outputs \citep{LeBouquin-2004b}.

But to measure the coherence factor of each baseline individually, the geometry of the multiaxial combiner has to be carefully checked. On the detector each pair of pupils produces a fringe pattern with a frequency given by their separation. In order to separate the energy of each fringe pattern in the Fourier plane, the output pupil configuration should be \emph{non-redundant}. This necessary condition can be achieved with bi-dimensional or linear arrangements. In the first approach, the focal image is fringed in different directions, while in the second one all fringes are aligned but use different frequencies. Only this last 1-D diffraction pattern can be injected into a slit of spectrograph, opening spectral abilities at medium and large resolutions. It explains why the majority of the projects only consider linearly aligned output pupils, as VITRUV \citep{Malbet-2004}, MIRC \citep{Monnier-2004}, MATISSE and VEGA (Mourard, private communication).

From Monte-Carlo simulations, \citet{Ribak-1988} derived geometries allowing up to 30 beams to be combined without any redundancy. However, the number of frequencies dramatically increases with the number of input beams. For instance, 35 different frequencies are required for 8 beams. Reducing this number has been initially proposed by \citet{Vakili-1989}, for a visible interferometer in the presence of a fully turbulent image (dispersed fringed speckles). The authors used a \emph{completely redundant} configuration but introduced small optical path differences to differently tilt the fringes in the dispersed image, and thus separate the information. However, to save the spectral resolution abilities, the spectral dimension should be oversampled and the total number of pixels remains the same: the coding is converted from spatial to spectro-spatial. This method has never been used but could probably be explored to combine a large number of telescopes. Nevertheless it is clearly out of the scope of this paper, since it requires a global analysis taking into account its technical specificity (high spectral resolution, spectro-spatial coding). To fit with existing and incoming instrumentation, our work focuses on \emph{partially redundant} output pupils configuration, without using an optical path difference.

In this case, to be able to recover the interferometric quantities, the fringed image pattern should not be blurred by the atmospheric turbulence. In other words, the image should be stabilized and not contain moving speckles. Practically, the common way to transform a corrugated input wavefront into a planar and stable output is to spatially filter the beams with single-mode fibers. It also drastically improves the accuracy of the instrument as demonstrated by the FLUOR experiment \citep{Perrin-1997,Foresto-1998} and theoretical studies \citep{Tatulli-2004}. Yet, the average phase difference between the pupils (the so-called piston) should remain constant over the exposure time. This is ensured by freezing the random atmospheric piston with short exposures of few millisecond, or by the help of an external fringe tracker that allows longer exposures. However, this is not an issue for our study, since the same frame-to-frame data processing can be applied to both short or long exposures. Finally, the average value (distance to white fringe) and the residual motion (high frequency jitter) of the piston during each exposure will lead to contrast losses that have to be calibrated. Again, this pure multiplicative factor on the fringe amplitude has no incidence in our study.

All these advantages explain why \emph{partially redundant multiaxial combination} and \emph{spatial filtering} is the setup currently used in the AMBER interferometer. And also why it is a contemplated solution for next-generation instrument. Even so, no studies of the influence of the pupil redundancy on the interferometric quantities and on their estimators have been published. The objective of this work is to fill this lack.
In Section~\ref{sec:multi}, we present the description of the single-mode all-in-one multiaxial combination. We clarify the relation between the pupil and the interferogram planes. We explain why and how to reduce the maximum coding frequency of the fringes by compressing the pupil plane. We recall four different estimators of the squared visibility that can be used in the case of single mode interferometry. In Section~\ref{sec:experiment} these estimators are compared in terms of relative performance. We investigate how they are robust with respect to the compression process. Finally section~\ref{sec:8beams} contains applications to incoming project making use of multiaxial all-in-one scheme with an increasing number of recombined beam. Thanks to results derived in previous sections, we propose an optimal output pupil configuration for 6- and 8-beam combiners.

This work is the second part of our study of single-mode multiaxial combination for astronomical interferometry. The work presented here makes an intensive use of the formalism and the results presented in the first paper \citep[paper~I, ][]{Tatulli-2005}.

\section{Single mode multiaxial combination} \label{sec:multi}

\label{sec:formalism}

\begin{figure}
  \begin{center}
    \includegraphics[scale=.45]{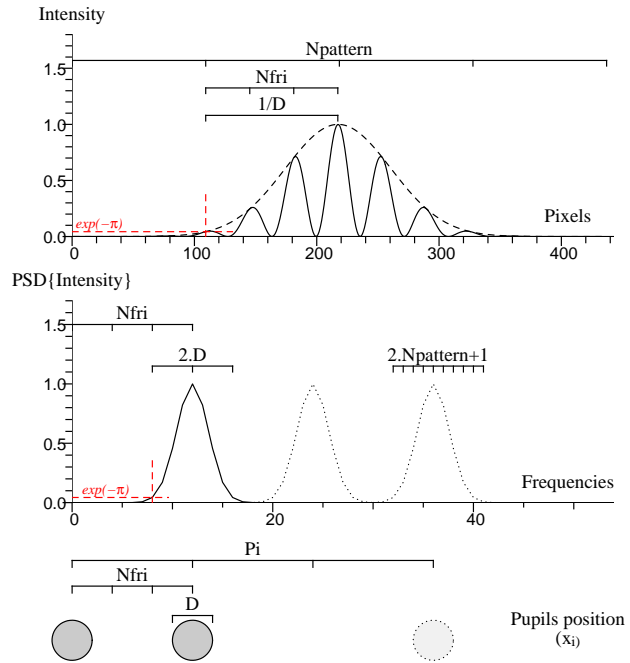}
    \caption{\label{fig:geometry} Geometrical relations between the image in the focal plane (top), its power spectral density (PSD, middle), and the output pupil configuration (bottom) of a multiaxial all-in-one single mode interferometer. Here is shown a 3-beam combiner when beam 1 and 2 are illuminated with an instrumental contrast set to 1. When beam 3 is illuminated, two other fringe systems appear (dotted lines in the Fourier space; the corresponding fringes in the image space have not been drawn for a sake of clarity). The displayed geometry is defined by $N_{fri}=3$, $N_{pattern}=4$ and $P_i=(0,1,3)$. See text for a complete explanation of these parameters.
}
  \end{center}
\end{figure}

We redefine here only the parameters relevant for this part of the study. For a detailed description of the single-mode multiaxial combination, the reader can refer to Section~2 of paper~I.
The different spatially filtered beams are surimposed in a focal plane, forming a fringed image. The amplitude and phase of this modulation, also called complex visibility, are related to the source intensity distribution at high angular resolution, by the Zernike van Cittert theorem. Strictly speaking the visibility obtained with a fibered interferometer is not the source visibility but the so-called modal visibility \citep{Mege-thesis}, that we will consider to be our observable.

%
%
%

Due to the spatial filtering, the shape of the image pattern and the fringe frequencies are completely defined by the width and the separation of the output pupils (Fig.~\ref{fig:geometry}). We called $D$ the width of the beams in a pupil plane, here defined by the extension of the fiber mode. The fringes are weighted by the diffraction pattern of the fiber mode in the image plane (here Gaussian, dashed line in Fig.\ref{fig:geometry} top). The number of fringes in this pattern ($N_{fri}$) is fixed by the distance between the output pupils. $N_{pattern}$ is the detector reading window size expressed in unit of the diffraction pattern, that is $N_{pattern}=1$ means half of the lobe of the Gaussian pattern has been considered. It also correspond to the number of independent frequency points under each peak (Fourier sampling law). Additionally, $P_i$ is a non-redundant integer list which fixes the relative positions of the other pupils and thus of the others peaks. Note that the largest frequency used in the Fourier plane identifies with the largest distance between two pupils. Thus, the position $x_i$ of the pupils $i$ (starting with $i=0$ for the first one) can be expressed as:
\begin{equation} \label{eq:pupils}
  x_i = P_i\;.\;N_{fri}\;.\;D 
\end{equation}
where $P_i$ is the non-redundant integer list:
\begin{equation} \label{eq:3pupils}
  P_i = (0,1,3) 
\end{equation}
\begin{equation} \label{eq:4pupils}
  P_i = (0,1,4,6)
\end{equation}
for a 3-beam (Eq.~\ref{eq:3pupils}) and a 4-beam combiner (Eq.~\ref{eq:4pupils}). With such assumptions, the peaks corresponding to each fringe pattern are equally spaced in the Fourier plane (Fig.~\ref{fig:geometry}, middle).

\subsection{Pupil plane compression} \label{sec:superposition} \label{sec:how} \label{sec:why}

To separate the Fourier peaks, the distance between the pupils should be a least twice the pupils width ($N_{fri}\ge2$). When increasing the number of beam, the non-redundant configuration $P_i$ reaches wide frequency ranges, such as 35 for a 8-beam combiner. Thus the largest distance between two pupils is $2\times35=70$ times the pupil width (Eq.~\ref{eq:pupils}). Using $N_{pattern}\ge2$ to record the whole image pattern, the number of fringes is $2\times35\times2=140$, and it requires about $2\times35\times2\times4=560$ pixels to sample them. A the same time, such combiner requires a minimal spectral resolution of about $600$, to be sure that the number of fringes in the coherence lenght is larger than the number of recorded fringes.
First, reducing the frequency range reduce the number of pixel used to code the signal, and thus the detector noise contribution. Besides, reducing the number of recorded fringe allows to reduce the minimal spectral resolution, when the dispersion is not mandatory by the science case.

\begin{figure}
  \begin{center}
    \includegraphics[scale=.40]{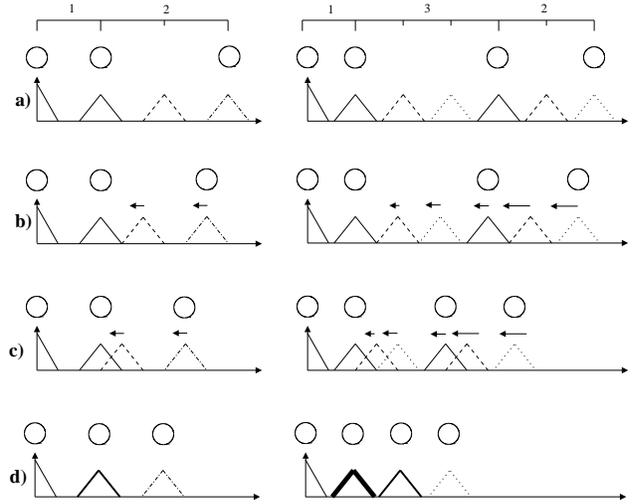}
    \caption{\label{fig:superposition} Superposition of the peak due to a compression of the pupil plane by Eq.~\ref{eq:pupils_conf} for a 3- and a 4-beam combiner. The Compression Factor is set to $\rho=1$, $2/N_{fri}$, $1/N_{fri}$ and $0$  (\textbf{a}, \textbf{b}, \textbf{c} and \textbf{d}).}
  \end{center}
\end{figure}

The first way to scale down the frequency range is to reduce $N_{fri}$. It corresponds to a homothety of the pupil and the Fourier planes without changing the pupil diameter and the peak width. Peaks start to overlap when $N_{fri}<2$, and the overlapping rate is similar for all of them. However, it is impossible to have $N_{fri}<1$ since the pupils cannot spatially overlap. As a result this method has limited compression capability. Another solution to perform more efficient compression is to keep the distance between the first and second pupils fixed (i.e the $N_{fri}$ parameter) and only rescale the position of third and further pupils. If we conserve an homothetic scaling of these latter -- namely the compression factor $\rho$ -- to keep a partial non-redundancy, the pupils positions now follow:
\begin{equation} \label{eq:pupils_conf}
x_i = \;(i + (P_i - i).\rho)\;\;.\,N_{fri}\,.\,D
\end{equation}
The smallest frequency does not shift, and the Fourier plane transformation is not perfectly homothetic. Gaps are created with a size of $N_{fri}$, the original space unit between the pupils. Interesting values are displayed on Figure~\ref{fig:superposition}~:
\begin{description}
\item[\textbf{a}~:] the pupils and Fourier peaks are at the positions defined by Eq.~\ref{eq:pupils}.
\item[\textbf{a}~$\to$~\textbf{b}~:] the Fourier peaks do not overlap each other, the configuration is not compressed.
\item[\textbf{b}~$\to$~\textbf{c}~:] the Fourier peaks overlap each other less than a half, the configuration is \emph{slightly compressed}.
 \item[\textbf{c}~$\to$~\textbf{d}~:] the Fourier peaks overlap more than a half of the peak width, the configuration is \emph{strongly compressed}.
 \item[\textbf{d}~:] the pupil space is constant, the configuration is fully redundant. In the Fourier plane, the peaks are superposed by group separated by $N_{fri}$.
\end{description}
Note that this method allows the Fourier plane to be compressed, even if it is impossible to juxtapose the closest pupils, for technical reasons for instance.

\subsection{Visibility estimators} \label{sec:estimators}

There are different ways to recover the individual visibility of each baseline from the recorded image. Before the advent of single mode interferometry, and to overcome the problem of turbulence, \citet{Roddier-1984b} developed estimators based on Power Spectral Density integration. In paper~I, we exposed the formal expressions of expected values and Signal to Noise Ratio (SNR) for this kind of estimators in the framework of filtered multiaxial combiners. In this paper, this method will be called \emph{PSD-Integration} (Integ).

It is also possible to measure the visibility by considering only the maximum value of the PSD of each peak. The required calibration is the same as for PSD-Integration. This estimator was not specifically studied in the previous paper, but it can be calculated with the same formalism with integration limited to only one pixel (the maximum) of the Fourier peak. This method will be called \emph{PSD-Maximum} (Max).

By taking more benefit from the deterministic properties of the PSD, the square visibility can be recovered by fitting the PSD shape. The shape has to be previously calibrated by observations of unresolved star or internal light. This estimator was not presented in our previous study. We implemented its theoretical expressions for expected values and SNR and checked them thanks to numerical experiments described in paper~I. This method will be called \emph{PSD-Fitting} (Fit).

\begin{figure}
  \begin{center}
    \includegraphics[width=.98\linewidth]
    {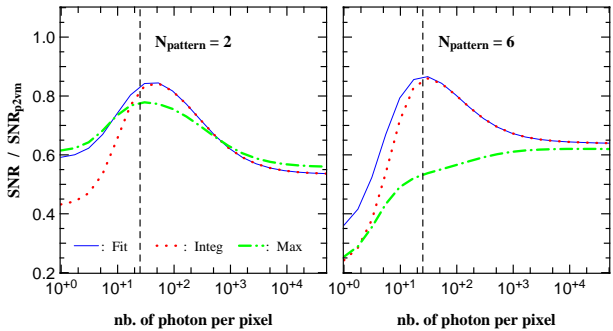}
    \caption{\label{fig:SNR_comp} Ratio between the SNR of the PSD-based estimators and the SNR of the model-based one, as a function of the number of photon per pixel in the interferogram. The source visibility is set to $\mu=0.5$. Detector noise is $\sigma=15$e$^-$/pix. The Fourier peaks are fully separated ($N_{fri}=2.5$ and no compression). The curves are plotted for a narrow and a large detector reading window (left and right).} 
  \end{center}
\end{figure}

\begin{figure*}
  \begin{center}
    \includegraphics[angle=-90,width=0.98\linewidth]
    {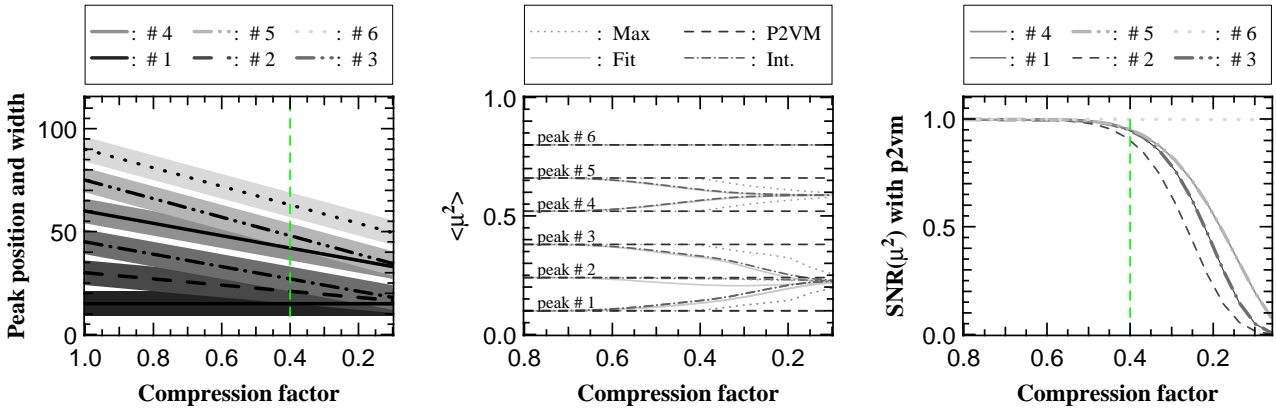}
   \caption{\label{fig:34T_p2vm_unbiased} \label{fig:34T_redundance} {\bf Left~:} Positions (thick lines) and supports (filled regions) of the fringe peaks in the Fourier space versus the compression factor. {\bf Middle~:} Square visibility ($\mu^2$) obtained with the different estimators described in Sec.~\ref{sec:estimators}. {\bf Right~:} SNR on the square visibility of each baseline computed for the model-based estimator normalized with respect to the zero-compression factor value ($\rho=1$). The vertical dashed line is for $\rho=1/N_{fri}=0.4$.}
  \end{center}
\end{figure*}

Since the shape of the interferogram is completely determinist, the visibilities can also be recovered by fitting the data. The fit can be identically done in the direct plane (fringe fit) or in the Fourier plane (complex Fourier transform fit). It can be formally expressed as a matricial relation between the recorded values on the detector and the complex visibility of each baseline: the so-called \emph{Pixel To Visibility Matrix} (P2VM). The visibility amplitudes and phases are recovered in the same inversion process. If no fringe tracking unit is available, the phase quantity is stochastic from frame to frame and is meaningless. It is ignored by averaging the square visibility instead of the complex one. As a main difference with the PSD-based algorithms presented above (Integ, Max and Fit), this square operation is done \emph{after} the information of the baselines have been separated. The calibration requires to record the instrumental fringe shapes with a good accuracy. The reader can refer to the AMBER data processing method for which such a model-based algorithm is currently used \citep{Millour-2004}.

\section{Numerical experiment}
\label{sec:experiment}

As far as an astronomer is concerned, the properties of interest when dealing with different estimators are the relative accuracy performance, the relative bias on the expected value and their dependencies with the instrumental parameters.

\subsection{Estimator relative performances without compression}
\label{sec:snr_comp}

Figure~\ref{fig:SNR_comp} illustrates the performance of the PSD-based estimators compared to the model-based one in different noise regimes, without compression, and for two different detector reading windows ($N_{pattern}$). The source visibility is arbitrarily set to $\mu=0.5$ and the detector noise is $\sigma=15$e$^-$/pix. The model-based estimator always presents better SNR, especially in the strong photon-rich and photon-poor domains. The PSD-based estimators follow more or less the same curves for a small detector reading window. However, when increasing the number of independent points over the fringe frequencies ($N_{pattern}>2$), the PSD-maximum shows worse performance, because it does not take into account the whole available information. At least, they all reach a similar asymptote in photon-noise limit, from where their SNR is about $0.6$ times the model-based one. This asymptote does not depend on the instrumental configuration but only function on the source visibility.

In paper~I, we have mentioned the advantage of model-based estimator versus PSD-integration, especially for unresolved sources with high visibility. Showing the computations presented here we conclude that this advantage can be extended over all the PSD-based estimators without exception. It tends to prove that a first-order estimation systematically drive a better -- or at worst identical -- performance than quadratic one.

\subsection{Estimator robustness to compression} \label{sec:bias}

To compare the estimator relative robustness to compression, we compute the recovered visibility amplitudes for a large range of compression factor. The initial setup is the same than in Sec.~\ref{sec:snr_comp} (4-beams, $N_{fri}=2.5$, $N_{pattern}=6$, and pupil positions given by Eq.~\ref{eq:4pupils}). For the sake of the clarity of Figure~\ref{fig:34T_p2vm_unbiased}, the input visibilities have been chosen arbitrarily different, and all the phases are set to zero. It has absolutely no incidence on the results, that have been validated with various set of input parameters (combiner geometry, fringe visibilities and phases).
Results are displayed in Figure~\ref{fig:34T_p2vm_unbiased}, with a compression factor ranging from $1$ to $0.1$. As soon as peaks overlap, PSD-integration and PSD-fit rapidly fail to recover to visibility. PSD-maximum only fails when peaks strongly overlap ($\rho\le 0.4$). Besides, the model-based estimation is never biased even when the compression factor reaches $0.1$.

The bias in the estimated visibility can be explained by mixing of information between corresponding peaks when they start to overlap. This blend is due to the coherent sum of the two interferograms under the overlapped frequencies. Because the input phases and optical path delays are arbitrarily zero, the asymptotic square visibility recovered is the square average of the two corresponding ``single'' visibilities, as shown in the right asymptotes of Fig.~\ref{fig:34T_p2vm_unbiased}. In a general way, it depends on both the visibilities and phases. Such a complex sum prevents from a simple calibration of this effect. The only solution is to previously calibrate the complex shape of each individual peak and then to inverse both the amplitudes and the phases at the same time. Doing-so one has just re-invented the model-based estimator in the Fourier space, and it explains why this last estimator is never biased.

\subsection{Signal-to-Noise Ratio versus compression} \label{sec:p2vm_snr}

We now focus on the model-based estimator since we have shown that this latter is the only one able to retrieve unbiased interferometric quantities even if the pupil plane is compressed. Formal expressions of the SNR computed in paper~I remain valid for all compression factors. Computations give similar results in both photon noise and detector noise regime. We present here only photon-rich computations. The instrumental setup used is the same as in Sec.~\ref{sec:bias}. Remember that it has been chosen for illustration purpose only and that we checked the results with various combiner configurations.

Right sub-panels of Fig.~\ref{fig:34T_redundance} shows the results, after normalization of the SNR by the value obtained without compression. Three different slopes can be distinguished~:
\begin{itemize}
\item Peak \#6 never overlaps. Its SNR remains constant.
\item Peaks \#4 and \#5 interact with one neighbor. Their SNR keep unchanged when peaks slightly overlap but rapidly decrease when peaks strongly overlap.
\item Peak \#2 interact with its two neighbors. Its SNR keeps unchanged for a slight overlap but presents the more abrupt decrease when peaks strongly overlap.
\end{itemize}
Slopes related to peaks \#1 and \#3 can be explained as a combination of the latter. They start with 2-peak interactions (each of them with a side of peak \#2). So their SNR follow the curve of peaks \#4 and \#5. Then they meet together over peak \#2 and begin also to be engaged in a 3-peak interaction. Their SNR reach the curve related to peak \#2.

The expression of model-based estimator as a matrix can help to understand these results. Each complex visibility (one per baseline) corresponds to one line of this matrix. When peaks overlap, corresponding lines begin more and more similar, and thus singular, which mathematically reduces the accuracy of the inversion. From this study, we conclude that -- whatever the initial peak position and the number of neighbors -- the SNR is unaffected by a slight overlap but goes down rapidly for a strong one. So compression factor of $1/N_{fri}$ can be applied to any configuration without damaging the performance, whatever the initial $N_{fri}$ parameter and the noise regime.

\subsection{Remarks}
In this section, the presented computations have been done with a Gaussian interferogram shape, which is the Fourier Transform of a beam filtered by a Gaussian mode. Strictly speaking, a fiber mode is purely Gaussian only if the profile of the refraction index is a Gaussian function too. However, it gives a good approximation in general with step-index fibers or waveguides. All the results have been also checked with Bessel interferogram envelop (corresponding to circular pupils), and lead to same conclusions. However, one should remember that turbulent beams (not perfectly corrected by adaptive optics or not spatially filtered) will lead to fringes that cannot be fitted by a first-order estimator.
Besides, the total number of pixels remains constant in our computations, although it becomes possible to reduce it when the Fourier plane is compressed (Shannon sampling is relaxed). We have tried to optimize the sampling rate for each compression factor, but it does not change the results.

\section{Application to 6- and 8- beam combiners}
\label{sec:8beams}
\begin{figure*}
  \begin{center}
    \includegraphics[angle=-90,width=0.98\linewidth]
    {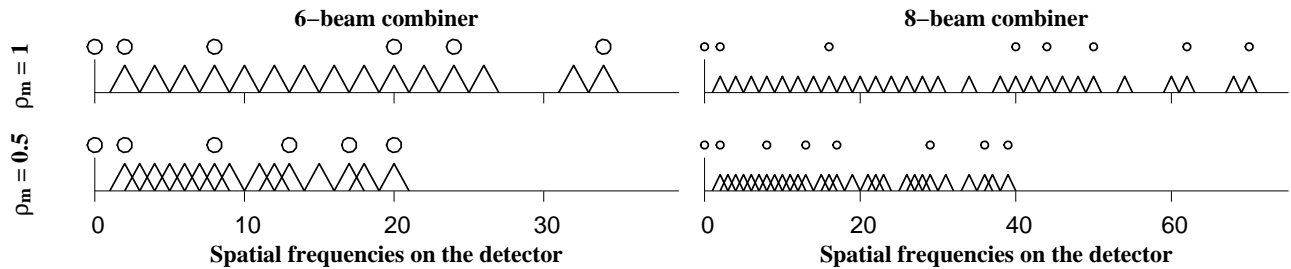}
    \caption{\label{fig:6-8-beam} Results of plane optimization for a 6-beam (left) and 8-beam (right) combiner. The output pupil configuration and the Fourier plane are represented at the same scale (autocorrelation relation). The pupil positions are computed for a Compression Factor of $\rho=1$ (top) and $\rho=0.5$ (bottom). The minimum distance between the pupils is set to twice the pupils width ($N_{fri}=2$). The frequency range is reduced by a factor $0.58$ and $0.55$ in the 6- and 8-beam case respectively.}
  \end{center}
\end{figure*}

For more than 4 beams, it is impossible to theoretically determine the best non-redundant pupil configuration $P_i$ which minimizes the maximum frequency used. This well-known problem can only be solved by ``brute force'' methods, and non-redundant integer lists have been obtained for a large number of telescopes \citep{Ribak-1988}, such as~:
\begin{equation} \label{eq:6_beams}
  P_i = (0,1,4,10,12,17)
\end{equation}
\begin{equation} \label{eq:8_beams}
  P_i = (0,1,8,20,22,25,31,35)
\end{equation}
for 6 and 8 beam respectively. The maximum of $P_i$ is larger than the number of baselines, even if the configuration is optimized. As a consequence, it leaves gaps in the frequency space. To compress the Fourier plane, we applied Eq.~\ref{eq:pupils_conf} on the 6-beam and 8-beam configurations. Surprisingly, some peaks fully overlap before the compression factor goes to zero. So we expected that some worse starting $P_i$ configuration (with more gaps) could be more compressed, and finally provide smaller frequency range.

To test it, we directly introduced the minimum allowed overlap rate in the ``brute force'' optimization algorithm as a new user-defined parameter ($\rho_{min}$). Fig.~\ref{fig:6-8-beam} shows the results with no compression (top, corresponding to Eq.~\ref{eq:6_beams} and \ref{eq:8_beams}) and with the maximum overlap rate allowed by our study (bottom, $\rho_{min}=1/N_{fri}$) for a 6- and 8-beam combiner. There is no simple transformation between the best pupil configurations with and without compression. We infer that it is impossible to found the best compressed configuration from the best un-compressed one. Empirically, we remark that the compact solution leaves more gaps in the frequency space which are over-compensated by the compression. As a consequence, the maximum frequency is not multiplied by a factor $\rho_{min}=1/N_{fri}=0.5$. We only reach a compression of $\sim0.58$ for the 6-beam combiner and $\sim0.55$ for the 8-beam combiner.

\section{Conclusion} \label{sec:discussion}

In this paper we have chosen to focus on multiaxial single mode interferometry concept because it is one of the most promising solution for future spectro-imaging interferometers with a large number of telescopes \citep{LeBouquin-2004b}. Also, it is currently used in the AMBER instrument. Based on this specific recombination scheme, we have proposed a simple method by compressing the output pupil plane with keeping a partial non-redundancy condition. We then have recalled the definition of four different visibility estimators that can be classified in two types: the estimators that make use of the Power Spectral Density of the interferogram and the estimator based on the model fitting of the interferogram.

First extending our analysis of paper~I, we have shown that without compression,the  model-based estimators drives to better performances than all PSD-based, whatever the noise regime (detector or photon noise) and the instrumental setup. This tends to prove that a first-order estimation systematically drive a better performance than quadratic one.
Then we have analyzed the effects of the compression on these estimators. We have demonstrated that a model-based estimator is as well the suitable algorithm to deal with compression. Indeed, by first separating the baseline complex visibilities before taking the modulus, the method prevents information mixing between the baselines that leads to bias the visibility. Yet, the compression reduce the Signal to Noise Ratio because the matrix used in the inversion process becomes more and more singular. However, the accuracy on the visibilities is not dramatically damaged before the peak maximum reaches the edge of its close neighbor, that we have called a slight overlap. 

As regards to the existing AMBER instrument, this study allows us to claim that the model-based estimator \citep[the so-called P2VM, ][]{Millour-2004} is the suitable algorithm, and that the pupil overlap rate used (slight overlap) only reduces the SNR by few percents.

As regards to future instruments dealing with a larger number of input beams, we propose to use this overlapping rate to minimize the required frequency range. Nonetheless, we have shown that the maximum frequency can be multiplied by a factor smaller than $0.6$. Such an optimization of the output pupil configuration has important consequences.
First, it reduces the number of pixel per spectral channel, leading to a smaller contribution of the detector noise and a better limiting magnitude.
Secondly it reduces the minimum required spectral resolution, and thus increases again the limiting magnitude when dispersion is not mandatory by the science case.
At the same time, by reducing the number of required pixels and/or spectral channels, it increases the reading speed, which is an important parameter if no fringe tracking unit is available.

From the technical point of view, Integrated Optics (IO) offers promising solutions to realize a multiaxial single mode combiner. This technology has been proved with both laboratory and sky experiments \citep{Berger-2001,LeBouquin-2004}, and multiaxial combiners have already been designed \citep{Berger-2000}. The size of the chip is directly related to the physical space between the output pupils. Since losses are mainly due to linear propagation in the waveguides, reducing the required distance between the beams, as presented in this work, will lead to a better global efficiency. The compactness of the planar optical component allows one to combine many beams in the same chip, which drastically reduces the instability and the required alignments. The observational strategies (number of baselines, wavelength...) can be adapted to the object thanks to the “plug and play” ability of IO combiners. Finally, output beams of the planar component can act as the input slit of a spectrograph, avoiding complex anamorphic optics.

\section*{Acknowledgments}

The authors want to warmly thank Florentin Milour, Karine Perraut, Jean-Philippe Berger and Fabien Malbet for their interesting remarks and helps. We are grateful for the valuable comments by the referee which helped to improve this article. All the calculations and graphics were performed with the freeware \texttt{Yorick}\footnote{\texttt{http://yorick.sourceforge.net/index.php}}.



\def\mnras{Mon. Not. of the Royal Astron. Soc.}
\def\aap{Astron. \& Astrophys.}
\def\aaps{Astronomy and Astrophysics, Supplement}
\def\apss{Astrophysics and Space Science}
\def\apjl{Astrophysical Journal, Letters}

\bsp
\label{lastpage}
\end{document}